# Handling Uncertainty in Social Lending Credit Risk Prediction with a Choquet Fuzzy Integral Model

Anahita Namvar, Mohsen Naderpour
Decision Systems and e-Service Intelligence Laboratory
Centre for Artificial Intelligence (CAI), Faculty of Engineering and IT
University of Technology Sydney (UTS)
PO Box 123, Broadway NSW 2007 Australia
Anahita.Namvar@gmail.com; Mohsen.Naderpour@uts.edu.au

*Abstract*— **As** *one of the main business models in the financial technology field, peer-to-peer (P2P) lending has disrupted traditional financial services by providing an online platform for lending money that has remarkably reduced financial costs. However, the inherent uncertainty in P2P loans can result in huge financial losses for P2P platforms. Therefore, accurate risk prediction is critical to the success of P2P lending platforms. Indeed, even a small improvement in credit risk prediction would be of benefit to P2P lending platforms. This paper proposes an innovative credit risk prediction framework that fuses base classifiers based on a Choquet fuzzy integral. Choquet integral fusion improves creditworthiness evaluations by synthesizing the prediction results of multiple classifiers and finding the largest consistency between outcomes among conflicting and consistent results. The proposed model was validated through experimental analysis on a real-world dataset from a well-known P2P lending marketplace. The empirical results indicate that the combination of multiple classifiers based on fuzzy Choquet integrals outperforms the best base classifiers used in credit risk prediction to date. In addition, the proposed methodology is superior to some conventional combination techniques.*

## I. INTRODUCTION

Peer-to-peer (P2P) lending, also known as social lending, has become more popular in recent years because it provides an online trading platform for lending money without the interference of traditional intermediaries, such as banks. Social lending companies reduce the cost of finance by connecting the lender and the borrower directly [1]. Despite the sophisticated mechanisms these platforms provide to evaluate credit risk, social lenders still face the risks associated with unsecured loans [2]. Perhaps more so, because most do not have enough expertise to extract risk knowledge from the information available. Moreover, from a profitability standpoint, recognizing default loans is not only critical for lenders but also for the sustainability of the P2P lending market. Therefore, correctly identifying credit risk to support decision making about the eligibility of a particular borrower has emerged as a critical problem for P2P lending platforms [1, 3].

P2P lending as a financial model has been studied extensively in recent years. Typically, credit risk evaluation in P2P lending involves statistical approaches and machine learning methods that aim to predict the creditworthiness of borrowers by considering loan evaluation as a binary classification problem [1, 4]. Most studies in this domain consider single classifiers, but the influence of those classifiers on each other has not yet been explored. Given that each classifier contains some uncertainty, combining classifiers should improve classification outcomes. Thus, this study presents a credit risk prediction model based on a Choquet fuzzy integral that combines three different classifiers.

Fuzzy integrals have been used extensively in classification problems, and their performance has been proven by many empirical studies [5, 6]. Moreover, various theoretical and experimental results show that a fuzzy integral fusion model not only improves classification accuracy but also extends a model's generalizability and recognition system robustness [5, 7]. Of the different fuzzy integral methods, Choquet fuzzy integrals are better suited to quantitative problems, such as classification by aggregating the information from multiple base classifiers that may agree or conflict with each other.

This paper makes several contributions to the literature in this new and fast-growing field of P2P lending. We proposed a novel fusion classification framework for credit risk prediction in social learning based on a Choquet integral. The proposed ensemble model for credit risk assessment not only improves base classifiers but also outperforms other fusion techniques. Moreover, adaptive fuzzy fusion is a theoretical advancement that we employed in this study to adjust the fuzzy density importance of each classifier based on its ability to recognize objects in certain types of classes.

The rest of the paper is organized as follows. Section 2 provides a literature review of the loan evaluation techniques in P2P lending markets and the Choquet fuzzy integral approach. Section 3 describes our research methodology. Section 4 presents the experimental results, and Section 5 provides our conclusions and future research directions.

## II. Literature Review

### A. Loan Evaluation in P2P Lending

P2P lending emerged as a new e-commerce platform in the financial marketplace and has brought new economic efficiencies to financing [8]. Compared to the abundant literature on loan evaluation for traditional banking institutes, there are a limited number of studies on credit risk prediction in P2P lending [1]. Emekter et al. [9] investigated the P2P loan characteristics for credit risk of borrowers by applying logistic regression and survival analysis. They find that internal ratings, debt-to-income ratios (DTI), FICO scores, and revolving line utilization are all highly associated with loan defaults. Guo et al. [8] developed an instance-based loan evaluation model to inform investment decision making in P2P lending. Malekipirbazari and Aksakalli [3] used various machine learning methods, such as random forest, logistic regression, k-nearest neighbor, and support vector machines, for loan classification purposes. They find that a random forest classifier outperforms other the classifiers when predicting a loan's status. They also find a random forest classifier is more effective than relying on existing financial metrics, like FICO or LC grades, which the Lending Club provides to help lenders make loan investment decisions. Xia et al. [1] employ cost-sensitive learning and extreme gradient boosting to develop a cost-sensitive boosted tree loan evaluation model to predict the creditworthiness of borrowers. In addition to some studies that focus on hard information, Ge et al. [10] address soft information, which is not directly related to a borrower's financial status or creditworthiness. They apply social network information to predict defaults. Serrano-Cinca and Gutiérrez-Nieto [11] developed a profit scoring model by applying a decision-tree-based classifier to predict the expected profitability of investing in P2P loans. Wang et al. [12] proposed an ensemble mixture random forest model based on survival analysis to predict the probability of defaults over time. Byanjankar et al. [2] consider neural networks as means to classifying loan applicants into default and non-default groups. Namvar et al. [13] developed a credit risk prediction framework that compares different resampling approaches in combination with outstanding classifiers. They demonstrate that random under-sampling and random forest classifiers are an efficient combination of classifier and resampling strategy for credit risk prediction in P2P lending.

### B. Fuzzy Integral

The outputs of base classifiers are usually imprecise or uncertain, and fuzzy integrals are one of the most computationally efficient approaches for handling these uncertainty issues [5]. Sugeno and Choquet integrals are common fuzzy integral approaches that have been used in different areas of mathematics, economics, pattern recognition, and machine learning [14]. Although both are considered by scholars as relevant fuzzy integral approaches, Choquet integrals have been widely extended and used in more disciplines than Sugeno intervals [15]. Hence, Choquet integrals have emerged as an efficient approach for information fusion and aggregating multiple classifier models [14]. A Choquet integral is a method of aggregation that represents both the importance of a classifier and its interactions with other classifiers [7]. It relies on the concept of fuzzy measures first introduced by Sugeno [16]. The definitions of fuzzy measures and the Choquet integral are as follows [17]:

Let $X$ be a set of classifiers, and let $P(X)$ denote the power of each set of $X$.

**Definition 1:** A discrete fuzzy measure of $X$ is a set function $g: P(X) \rightarrow [0,1]$ that satisfies the following conditions:
1) Boundary conditions: $g(\phi) = 0, g(X) = 1$
2) Monotonicity:
   If $A, B \in P(X)$ and $A \subset B$ then $g(A) \leq g(B)$

$g(S)$ is interpreted as the grade of subjective importance of the classifier set $S$. The fuzzy measure values of the singletons, $g(x_i) = g^i$ are commonly called the densities. In addition to the worth of singletons, the worth of any combination of classifiers must also be calculated. To calculate the fuzzy measure of any combination of classifiers, the Sugeno λ-measure is used. This measure is defined by the values of the fuzzy densities. The λ-measure has the following additional property.

$$\begin{cases} g_\lambda(A \cup B) = g_\lambda(A) + g_\lambda(B) + \lambda g_\lambda(A) g_\lambda(B) \\ \forall A, B \in P(X), A \cap B = \phi \end{cases} \quad (1)$$

where $\lambda$ can be found by solving Equation 2.
$$\lambda + 1 = \prod_{i=1}^{n}(1 + \lambda g^i), \lambda > -1 \quad (2)$$

Therefore, fuzzy measures provide a way of quantifying the worth of combinations of classifiers based on Equation 1.

**Definition 2:** Let $g$ be a fuzzy measure of $X = \{x_1, x_2, \ldots, x_n\}$. The Choquet integral of function $f: X \to R$ with respect to $g$ is shown in Equation 3.
$$C_g(f) = \sum_{i=1}^{n} f_i [g(A_i) - g(A_{i-1})] \quad (3)$$
where ($i$) indicates a permutation of $X$ such that $f(x_{(1)}) \leq f(x_{(2)}) \leq \cdots \leq f(x_{(n)})$ also $A_i = \{x_{(i)}, x_{(i+1)}, \ldots, x_{(n)}\}$, $A_0 = \phi$.

$f_i$ denotes the prediction results of classifier $x_i$, $[g(A_i) - g(A_{i-1})]$ indicates the relative importance of the classifier $x_i$. The fuzzy integral of $f$ with respect to $g$ is the integration result.

### III. RESEARCH METHODOLOGY

In this research, we propose an ensemble classifier based on a Choquet fuzzy integral to improve credit risk prediction in P2P lending. A Choquet fuzzy integral can improve classification performance by capturing the usual interactions between base classifiers during the aggregation process. Fig. 1 illustrates this model.

#### A. Pre-processing

Data pre-processing is a crucial step before building a model. The main purpose of this step is to improve data reliability by cleaning the data, removing the outliers, eliminating null values, selecting the appropriate features, and transforming the data. Data pre-processing begins with data cleansing, which removes outliers, missing values, and null values. In feature selection, leakage fields are removed and appropriate features are selected using correlation analysis. During data transformation, categorical features are converted into numeric values, and standardization and log transformations may also be applied. Finally, since the training dataset is imbalanced, a resampling approach is applied to balance the dataset. In our approach, we followed the data pre-processing procedure described by Namvar et al. [13] .The output of this procedure is a clean and balanced dataset that is ready for analytics.

#### B. Base Classifiers

Three classifiers were selected and trained for the purposes of this study: a gradient boosting classifier, an adaboost classifier, and a logistics regression classifier. Gradient boosting is a highly effective classification algorithm that is commonly used in risk assessment for predicting the probability of default. Brown and Mues [18] show that gradient boosting is an effective classification technique for credit scoring, specifically where a class imbalance problem exists in the dataset. AdaBoost is one of the most popular classification methods. With this method, the final classifiers are derived by combining weak learners [19]. AdaBoost also performs well with class-imbalanced data [20]. Logistics regression has been widely used in credit scoring and is known as the industry standard algorithm for solving classification and regression problems. The output of this step is trained base classifiers.

#### C. Preliminary Fuzzy Densities

According to Equation 3, the Choquet fuzzy integral operates on the fuzzy measures (g), which are calculated based on the degree of importance of each classifier or the degree of importance of a subset of classifiers. Moreover, according to Equation 1, to calculate the worth of any combination of classifiers, the fuzzy density, also known as a singleton, should be calculated first. This parameter is critical and presents a difficult point in the practical application of fuzzy integral fusion. In this paper, the fuzzy densities of the basic classifiers are determined according to the confusion matrix of each classifier.

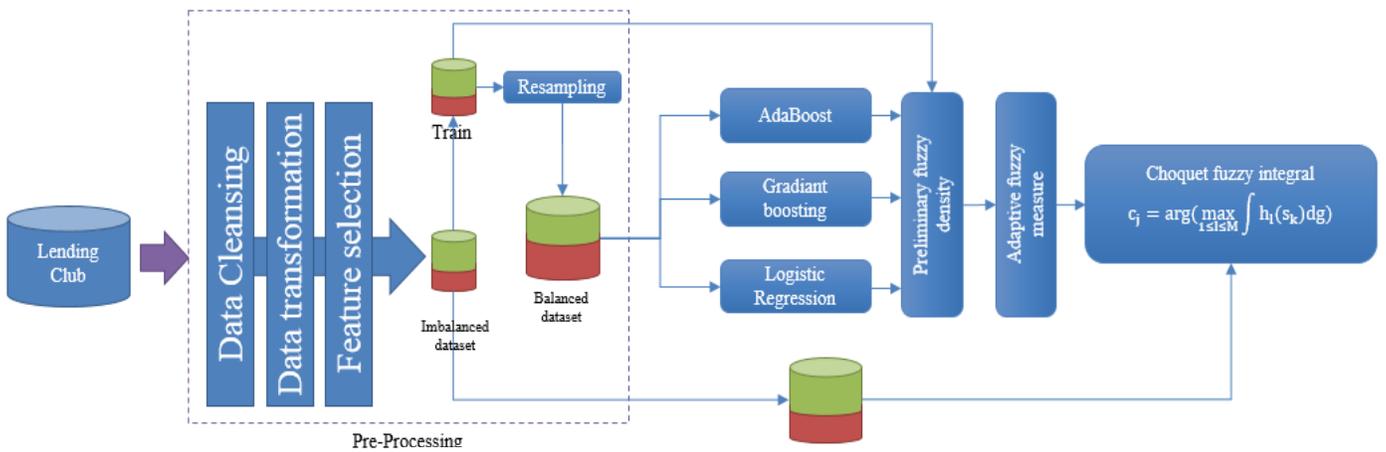

Fig. 1. The Choquet fuzzy integral prediction model.

Suppose the confusion matrix of the i-th classifier is defined as $CM_i = (n^i_{j_1 j_2})$:

$$CM_i = \begin{bmatrix} n^i_{11} & \cdots & n^i_{1M} \\ \vdots & \ddots & \vdots \\ n^i_{M1} & \cdots & n^i_{MM} \end{bmatrix} \quad i = 1, 2, \dots, P \quad (4)$$

where $j_1 = j_2$, $n^i_{j_1 j_2}$ represents the number of samples that belong to class $c_{j1}$ and is correctly classified as $c_{j1}$ by the i-th classifier. In the condition that $j_1 \neq j_2$, $n^i_{j_1 j_2}$ represents the number of samples that belong to class $c_{j1}$ but have been misclassified as $c_{j2}$ by the classifier i.

Therefore, given the precondition that the i-th classifier identifies sample $s_k$ as class $c_{j1}$, the conditional probability of this sample truly belonging to class $c_j$ is represented as follows:

$$p^i_{j1j} = p(s_k \in c_{j1} | E_i(s_k) = c_j) = \frac{n^i_{j1j}}{\sum_{j=1}^{M} n^i_{j1j}} \quad (5)$$

$$(j_1 = 1, 2, \dots, M; j = 1, 2, \dots, M)$$

Hence, the probability matrix is

$$PM_i = \begin{bmatrix} p^i_{11} & \cdots & p^i_{1M} \\ \vdots & \ddots & \vdots \\ p^i_{M1} & \cdots & p^i_{MM} \end{bmatrix} \quad i = 1, 2, \dots, P \quad (6)$$

The diagonal elements $p^i_{jj}$ in $PM_i$ represent the percentage of correct classifications by classifier $E_i$. Let $g^i_j = p^i_{jj}$, then $g^i_j$ represents the preliminary fuzzy density for the j-th class with respect to the i-th classifier. The output of this step is the fuzzy density of each base classifier for each of the different class labels.

D. *Adaptive Fuzzy Measures*

Given that each classifier may not perform equally as well in identifying all objects, i.e., one classifier may be more robust than others in recognizing certain types of classes but may be more error-prone in classifying other classes, assigning a fixed value to each fuzzy density tends to be an ineffective approach. Therefore, the fuzzy density $(g^i_j)$ should be properly adjusted according to each type of class and the information obtained by the other classifiers.

First, the fuzzy densities are updated by considering the proportion of the correct classifications and misclassifications within the classifier. Second, the fuzzy densities are further updated by considering the pairwise proportion of the misclassified objects between the considered classifier and the others. Thus, the fuzzy densities can be adjusted using the following parameters [21].

$$g^{*i}_j = g^i_j * \left( \prod_m \delta^{i/m}_j \right)^{w1} * \left( \prod_m \gamma^{i/m}_j \right)^{w2} \quad (7)$$

Where $g^{*i}_j$ is the updated fuzzy density of the classifier $E_i$ for class $C_j$; $\{\delta^{i/m}_j\}, 0 < \delta^{i/m}_j < 1$, and $\{\gamma^{i/m}_j\}, 0 < \gamma^{i/m}_j < 1$ are the sets of updated parameters. $g^{*i}_j$ is calculated for each of the classes. Each set of updated parameters may have a different effect on the classification output, so $w_1$ and $w_2$ add flexibility into the update processe. These two sets of the updated parameters are calculated as follows.

The term $\delta^{i/m}$ is used to update the initial fuzzy density in the sense that when the outputs of two classifiers do not agree with each other. Then, the initial fuzzy density of the considered classifier $(E_i)$ should be weakened in proportion to the number of misclassified objects, given the other classifier $(E_m)$ classified those objects correctly.

That is

$$\delta^{i/m}_j = f(x) = \begin{cases} 1 & , k_i/i = k_2/m \\ \dfrac{p^i_{j/i,j/i} - p^i_{k/i,j/m}}{p^i_{j/i,j/i}} & , k_1/i \neq k_2/m \end{cases} \quad (8)$$

where the symbols $k_1/i$, and $k_2/m$ indicate that class $k_1$ is given by classifier $E_i$, and class $k_2$ is given by classifier $E_m$, respectively. When $k_1/i = k_2/m$, this means that the two classifiers have identified a sample as belonging to similar classes. When $k_1/i \neq k_2/m$, the two classifiers have recognized a sample as belonging to different classes. In other words, $E_i$ has misclassified the object; however, the other classifiers, $E_m$, may have classified that object correctly. $P^i_{j/i,j/i}$ represents the proportion of objects correctly classified by the classifier $E_i$ for class $j$, and $p^i_{k/i,j/m}$ indicates the proportion of misclassified objects by classifier $E_i$ but correctly classified by other classifier $E_m$. Both $P^i_{j/i,j/i}$ and $p^i_{k/i,j/m}$ are obtained from the training dataset. The more objects misclassified by $E_i$ that are classified correctly by the classifier $E_m$, the more the corresponding fuzzy density of classifier $E_i$ for class $j$ is reduced.

The idea behind updating the parameters $\gamma^{i/m}_j$ is that if both classifiers make mistakes then the initial fuzzy density of a classifier should be reduced, but no changes to the fuzzy density value are required if the classifier $E_i$ makes the same or less mistakes than the classifier $E_m$. This is represented as

$$\gamma^{i/m}_j = \begin{cases} 1 & : p^i_{k/i,q/m} \leq p^m_{k/i,q/m} \\ \frac{p^m_{k/i,q/m}}{p^i_{k/i,q/m}} & : p^i_{k/i,q/m} \geq p^m_{k/i,q/m} \\ \varepsilon & : p^m_{k/i,q/m} = 0 \end{cases} \quad (9)$$

where $\varepsilon$ is given small real value, which prevents $\gamma^{i/m}_j$ from being zero.

The output of this step is an adjusted fuzzy density that updates the importance of each base classifier based on the classifier's performance in correctly classifying or misclassifying objects compared to the performance of other selected base classifiers.

*E. Choquet Fuzzy Integral*

According to [7], each single classifier has advantages and limitations. However, Choquet integral fusion can aggregate the prediction results of multiple classifiers and identify the results that share the greatest consistency.

Suppose the sample space of data $S$ can be divided into classes $C$ by classifiers $E$. Let $i$ be the classifier index ($i = 1, ..., P$) ; j be the class index ($j = 1, ..., M$); $k$ be the instance index ($k = 1, ..., N$). For the *k-th* substance, the prediction result output by the *i*-th classifier as $[h_{i1}(k), h_{i2}(k), ..., h_{iM}(k)]$ where $h_{ij}(k)$ represents the probability that the *i*-th classifier classified the *k-th* data to class $j$.

We have defined $[h_{1j}(k), h_{2j}(k), ..., h_{Pj}(k)]^T$ as $h_j(s_k)$ which can be interpreted as the function

$h_j: S \to [0,1]$ ,$h_j(s_k) = [h_{1j}(k), h_{2j}(k), ..., h_{Pj}(k)]^T$ Given a sample $s_k$, we can obtain a value for $h_j(s_k)$ that is defined as the degree of support given by each classifier with respect to the *j-th* class for sample $s_k$.

In addition to $h_j(s_k)$, the Choquet fuzzy integral operates on the fuzzy measures *(g)*. Fuzzy measures include fuzzy densities and the fuzzy measure of any combination of classifiers, which are calculated in Equations 8 and 1.

Then by calculating the Choquet integral of $h_j(s_k)$, $g$, we can provide the degree of support given by the ensemble classifier with respect to the *j-th* class for sample $s_k$. The output class $c_j$ for the sample $s_k$ is the class with the largest integral value:

$$c_j = arg(\max_{1 \leq l \leq M} \int h_l(s_k) dg) \quad (10)$$

A summary of the adaptive Choquet fuzzy integral algorithm follows.

---

1) **Construct the confusion matrix for each classifier as described by Equation 4, using the training dataset**
2) **For each j in [1… M] as class labels:**
   ✓ **For each i in [1…P] as classifiers:**
      - **Calculate the initial fuzzy densities defined by Equation 5**
      - **Calculate the updating parameter $[\delta^{i/m}_j]$ defined by Equation 8**
      - **Calculate the updating parameter $[\gamma^{i/m}_j]$ defined by Equation 9**
      - **Update the initial fuzzy densities using Equation 7**
      - **End for i**
   ✓ **Compute the $g_\lambda$- fuzzy measures using the updated fuzzy densities**
   ✓ **Compute the fuzzy integral defined by Equation 3 for each class**
   ✓ **End for j**
3) **Use Equation 10 to assign the sample to the output class**

---

IV. EXPERIMENTAL RESULTS

The purpose of these experiments is to examine the performance of the proposed fusion model based on the Choquet fuzzy integral for credit risk prediction in P2P lending. The Lending Club 2016-2017 dataset was used as the data, which is publicly available at Lendingclub.com. The Lending Club is the world's largest P2P lending platform. The dataset comprises 145 features spanning demographic, financial, and loan information for approximately 636,000 borrowers.

A binary classification problem with two class labels, "fully paid" and "default" loans, forms the basis of the experiments. Fully paid describes loans that are expected to be successfully paid back. Loans are classified as in default when a payment is later than 150 days past due. The data was pre-processed including data cleansing, feature selection, and data transformation, after which 66,376 records and 43 features remained. A brief description of the final data is provided in Table I.

TABLE I.  DATASET DESCRIPTION

| N | Features | % Default | % Fully paid |
|---|---|---|---|
| **66,376** | 43 | 18.3 | 81.7 |

The original dataset contains imbalances, so the data was resampled by under-sampling the majority class with randum under sampling (RUS) method .

In order to verify our choquet fuzzy integral classifier that we have proposed in this study, in this section we conduct a comparison analysis against three base classifiers: gradient boosting classifier, AdaBoost, and logistics regression. The balanced training dataset was used to train each model. The fuzzy densities were defined according to Equation 5 using a confusion matrix for each classifier, and the updated parameters were calculated based on Equations 8 and 9. Additionally, the optimal values of the exponential weights, w1,w2, were derived through a grid search analysis, which resulted in W1=0.9 , W2=0.6 as the best final pair. In all experiments, the value of ε in Equation 9 was set to 0.0001. The fuzzy densities were then adapted based on the updated parameters and the optimal values for W1 and W2.

The Sugeno λ-measures and, consequently, the worth of different sets of classifiers were calculated according to Equations 2 and 1, respectively, followed by the prediction results of the base classifiers, the $h_j(s_k)$ as discussed in Part E of Section 3. The ensemble model was built in Python.

To evaluate the model's performance, we conducted 5-fold cross-validation and averaged the results. Two performance measures were used to assess performance: AUC and G-mean. AUC reflects the area under the receiving operating curve (ROC). G-mean measures the tradeoff between specificity and sensitivity. Each measure has proven to be reliable in scenarios with class imbalanced issues [22].

The results of combining the three classifiers through a Choquet fuzzy integral approach compared to the performance of each base classifiers alone are shown in Table II.

TABLE II.  EXPERIMENTAL RESULTS

| | Methods | AUC | G-mean |
|---|---|---|---|
| Base classifiers | Gradient boosting | 70.54 | 64.52 |
| | Adaboost | 69.77 | 63.81 |
| | Logistic regression | 70.62 | 64.11 |
| Conventional combination methods | Optimistic OWA | 70.30 | 64.41 |
| | Pessimistic OWA | 70.17 | 60.39 |
| | Majority voting | 70.91 | 64.31 |
| Fuzzy integral combination | Proposed fusion model | **71.09** | **64.84** |

The results show that the Choquet integral-based combination classifier delivered better performance according to both performance measures. Hence, the results indicate that the aggregated model outperforms the base classifiers. Moreover, compared to existing fusion techniques, i.e., majority voting [23] and ordered weighting averaging (OWA) [24], the Choquet fuzzy integral performed better.

In terms of AUC, the proposed fusion model reached 71.09, which is higher than all the base classifiers, and also higher than the combination methods including OWA and majority voting. In terms of the G-mean measure, the proposed ensemble model resulted in 64.84 %, again higher than other classifiers.

To further verify the validity of the Choquet fuzzy integral-based combination model, a comparative analysis was conducted across all 5-fold cross-validation runs. Fig. 2 shows that the Choquet fuzzy integral had a higher AUC than the base classifiers, i.e., gradient boosting, Adaboost, and logistics regression, across all 5-fold cross-validation runs.

From Fig. 3, it can be seen that proposed ensemble model improved the G-Mean measure.

Overall, the experimental results show that aggregating multiple classifiers using a Choquet fuzzy integral can effectively improve the AUC and G-mean of a credit risk prediction model because this approach synthesizes the prediction results from multiple classifiers to reap the advantages of each classifier. Hence, our proposed combination method based on the Choquet fuzzy integral is more suitable than existing techniques for credit risk prediction in the P2P lending marketplace.

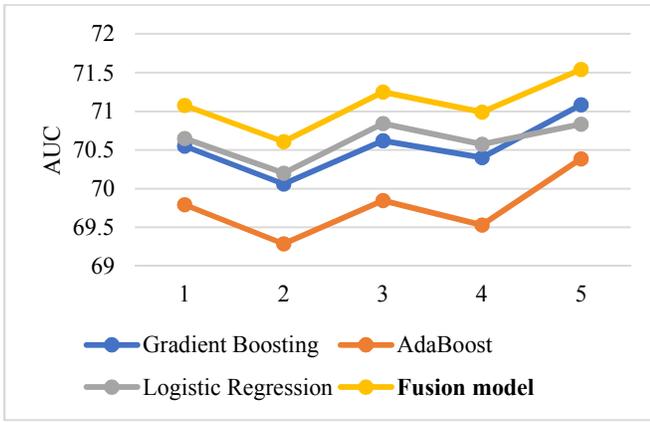

Fig. 2. AUC measurement of the Choquet integral fusion model and base classifiers on 5-fold cross-validation runs

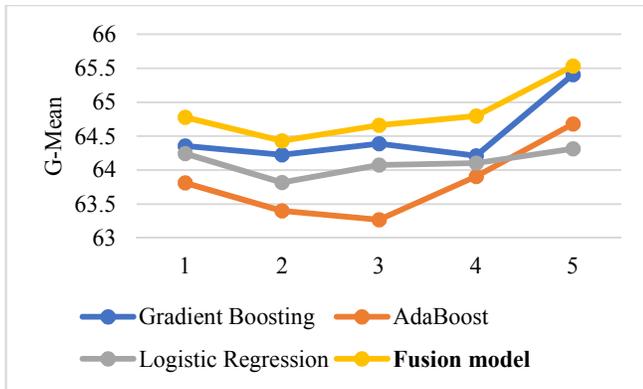

Fig. 3. G-Mean measure of the Choquet integral fusion model and base classifiers on 5-fold cross-validation runs

## V. Conclusion

Peer-to-peer lending is an innovative business model in financial technology that is bringing new opportunities for lenders and borrowers. However, although the emergence of social lending marketplaces reduces financial costs, the inherently insecure characteristics of P2P loans can result in huge financial losses for P2P platforms. Thus, this paper presents an ensemble method based on Choquet fuzzy integral to improve the accuracy of multiple classifiers in predicting the credit risk of borrowers in P2P lending. The proposed model synthesizes the prediction results of well-performing classifiers to take advantage of each single classifier in a way that considers both the redundancy and complementarity of the interactions between the outcomes of the classifiers. Moreover, through adaptive fuzzy measures, the model considers the differences in how well each classifier recognizes an object in each class. The experimental results demonstrate outstanding performance for the method in terms of AUC and G-mean measures, indicating that this ensemble model can increase the performance of credit risk predictions in P2P lending and support lenders in making credit risk decisions. In future work we want to improve adaptive fuzzy parameters to increase the performance of credit risk prediction model. Furthermore, we want to extend this study for dynamic credit risk prediction.

References

[1] Y. Xia, C. Liu, and N. Liu, "Cost-sensitive boosted tree for loan evaluation in peer-to-peer lending," Electronic Commerce Research and Applications, vol. 24, pp. 30-49, 2017.

[2] A. Byanjankar, M. Heikkilä, and J. Mezei, "Predicting credit risk in peer-to-peer lending: A neural network approach," in Computational Intelligence, 2015 IEEE Symposium Series on, 2015, pp. 719-725: IEEE.

[3] M. Malekipirbazari and V. Aksakalli, "Risk assessment in social lending via random forests," Expert Systems with Applications, vol. 42, no. 10, pp. 4621-4631, 2015.

[4] M. Siami, M. R. Gholamian, J. Basiri, and M. Fathian, "An Application of Locally Linear Model Tree Algorithm for Predictive Accuracy of Credit Scoring," in International Conference on Model and Data Engineering, 2011, pp. 133-142: Springer.

[5] X.-Z. Wang, R. Wang, H.-M. Feng, and H.-C. Wang, "A new approach to classifier fusion based on upper integral," IEEE transactions on cybernetics, vol. 44, no. 5, pp. 620-635, 2014.

[6] Y. Cao, "Aggregating multiple classification results using Choquet integral for financial distress early warning," Expert Systems With Applications, vol. 39, no. 2, pp. 1830-1836, 2012.

[7] X. Li, F. Wang, and X. Chen, "Support Vector Machine Ensemble Based on Choquet Integral for Financial Distress Prediction," International Journal of Pattern Recognition and Artificial Intelligence, vol. 29, no. 04, p. 1550016, 2015.

[8] Y. Guo, W. Zhou, C. Luo, C. Liu, and H. Xiong, "Instance-based credit risk assessment for investment decisions in P2P lending," European Journal of Operational Research, vol. 249, no. 2, pp. 417-426, 2016.

[9] R. Emekter, Y. Tu, B. Jirasakuldech, and M. Lu, "Evaluating credit risk and loan performance in online Peer-to-Peer (P2P) lending," Applied Economics, vol. 47, no. 1, pp. 54-70, 2015.

[10] R. Ge, J. Feng, B. Gu, and P. Zhang, "Predicting and deterring default with social media information in peer-to-peer lending," Journal of Management Information Systems, vol. 34, no. 2, pp. 401-424, 2017.

[11] C. Serrano-Cinca and B. Gutiérrez-Nieto, "The use of profit scoring as an alternative to credit scoring systems in peer-to-peer (P2P) lending," Decision Support Systems, vol. 89, pp. 113-122, 2016.

[12] Z. Wang, C. Jiang, Y. Ding, X. Lv, and Y. Liu, "A novel behavioral scoring model for estimating probability of default over time in Peer-to-Peer lending," Electronic Commerce Research and Applications, 2017.

[13] A. Namvar, M. Siami, F. Rabhi, and M. Naderpour, "Credit risk prediction in an imbalanced social lending environment," International Journal of Computational Intelligence Systems (OA), 2018.

[14] Q. Wang, C. Zheng, H. Yu, and D. Deng, "Integration of Heterogeneous Classifiers Based on Choquet Fuzzy Integral," in Intelligent Human-Machine Systems and Cybernetics (IHMSC), 2015 7th International Conference on, 2015, vol. 1, pp. 543-547: IEEE.

[15] A. R. Krishnan, M. M. Kasim, and E. M. N. E. A. Bakar, "A short survey on the usage of Choquet integral and its associated fuzzy measure in multiple attribute analysis," Procedia Computer Science, vol. 59, pp. 427-434, 2015.

[16] M. Sugeno, "Theory of fuzzy integrals and its applications," Doctorial Thesis, 1974.

[17] T. Murofushi and M. Sugeno, "An interpretation of fuzzy measures and the Choquet integral as an integral with respect to a fuzzy measure," Fuzzy sets and Systems, vol. 29, no. 2, pp. 201-227, 1989.

[18] I. Brown and C. Mues, "An experimental comparison of classification algorithms for imbalanced credit scoring data sets," Expert Systems with Applications, vol. 39, no. 3, pp. 3446-3453, 2012.

[19] S. Jadhav, H. He, and K. W. Jenkins, "An Academic Review: Applications of Data Mining Techniques in Finance Industry," International Journal of Soft Computing and Artificial Intelligence, vol. 4, no. 1, pp. 79-95, 2017.

[20] W.-C. Lin, C.-F. Tsai, Y.-H. Hu, and J.-S. Jhang, "Clustering-based undersampling in class-imbalanced data," Information Sciences, vol. 409, pp. 17-26, 2017.

[21] T. D. Pham, "Combination of multiple classifiers using adaptive fuzzy integral," in Artificial Intelligence Systems, 2002.(ICAIS 2002). 2002 IEEE International Conference on, 2002, pp. 50-55: IEEE.

[22] S. Wang, L. L. Minku, and X. Yao, "Online class imbalance learning and its applications in fault detection," International Journal of Computational Intelligence and Applications, vol. 12, no. 04, pp. 1340001.1-1340001.19, 2013.

[23] M. Siami, M. R. Gholamian, and J. Basiri, "An application of locally linear model tree algorithm with combination of feature selection in credit scoring," International Journal of Systems Science, vol. 45, no. 10, pp. 2213-2222, 2014.

[24] J. Basiri, F. Taghiyareh, and B. Moshiri, "A Hybrid Approach to Predict Churn," in Services Computing Conference (APSCC), 2010 IEEE Asia-Pacific, 2010, pp. 485-491: IEEE.